\documentclass[journal]{IEEEtran}
\IEEEoverridecommandlockouts
\usepackage{cite}
\usepackage{amsmath,amssymb,amsfonts}
\usepackage{graphicx}
\usepackage{textcomp}
\usepackage{xcolor}
\usepackage{algorithmic}
\usepackage{algorithm}
\usepackage{float}
\usepackage{flushend}

\title{RL-based Adaptive Task-Offloading in Mobile-Edge Computing for Future IoT Networks}

\author{
    Ziad Qais Al-Abbasi\IEEEauthorrefmark{1}, 
    Khaled M. Rabie\IEEEauthorrefmark{2},~\IEEEmembership{Senior Member,~IEEE},
    Xingwang Li\IEEEauthorrefmark{3},~\IEEEmembership{Senior Member,~IEEE}, \\
    Wali Ullah Khan\IEEEauthorrefmark{4},~\IEEEmembership{Senior Member,~IEEE}, and 
    Asma Abu Samah\IEEEauthorrefmark{5}
    
    \thanks{\IEEEauthorrefmark{1}Ziad Qais Al-Abbasi is at the Middle Technical University (MTU), Baquba Technical College (BTC), Baghdad 10074, Iraq. ORCID: 0000-0001-7062-3383 (Email: ziad.al-abbasi@mtu.edu.iq).}
    \thanks{\IEEEauthorrefmark{2}Khaled Rabie is with the Department of Computer Engineering, King Fahd University of Petroleum \& Minerals (KFUPM), Dhahran, Saudi Arabia.}
    \thanks{\IEEEauthorrefmark{3}Xingwang Li is with the School of Physics and Electronic Information Engineering, Henan Polytechnic University, China.}
    \thanks{\IEEEauthorrefmark{4}Wali Ullah Khan is with the Interdisciplinary Centre for Security, Reliability and Trust (SnT), University of Luxembourg, 1855 Luxembourg City, Luxembourg (Email: waliullah.khan@uni.lu).}
    \thanks{\IEEEauthorrefmark{5}Asma Abu Samah is with the Department of Electrical, Electronics and Systems Engineering, Universiti Kebangsaan Malaysia (Email: asma@ukm.edu.my).}

}
\begin{document}
\maketitle

\begin{abstract}
The Internet of Things (IoT) has been increasingly used in our everyday lives as well as in numerous industrial applications. However, due to limitations in computing and power capabilities, IoT devices need to send their respective tasks to cloud service stations that are usually located at far distances. Having to transmit data far distances introduces challenges for services that require low latency such as industrial control in factories and plants as well as artificial intelligence-assisted autonomous driving. To solve this issue, mobile edge computing (MEC) is deployed at the network's edge to reduce  transmission time. In this regard, this study proposes a new offloading scheme for MEC-assisted ultra-dense cellular networks using reinforcement learning (RL) techniques. The proposed scheme enables efficient resource allocation and dynamic offloading decisions based on varying network conditions and user demands. The RL algorithm learns from the network's historical data and adapts the offloading decisions to optimize the network's overall performance. Non-orthogonal multiple access is also adopted to improve resource utilization among the IoT devices. Simulation results demonstrate that the proposed scheme outperforms other state-of-the-art offloading algorithms in terms of energy efficiency, network throughput, and user satisfaction. 
\end{abstract}


%

\section{Introduction}
%
%
%
%
\IEEEPARstart{R}{ecently}, Internet of Things (IoT)-enabled 6G networks and  computing paradigms have gained a lot of attention in both industry and academia where Mobile
Edge Computing (MEC) has emerged with the goal of delivering cloud power at the network edge, addressing key issues such as connectivity, latency, and bandwidth that Cloudlet Computing (CC) cannot handle on its own \cite{ref1,ref2}. MEC, CC, and MCC are some of the implementations of edge computing principles that have been presented as a result.
One example to address this issue, MEC has emerged as a new concept \cite{ref3}, which brings computing and storage resources closer to the mobile network edge, allowing highly demanding IoT-based applications to run on devices while still maintaining the delay constraints \cite{ref4,ref5}. Another advantage of MEC confirming its resource-efficient policy is that its resources could also be exploited by other nearby entities such as operators and third parties for performing specific purposes. 
     
By incorporating intelligence into edge devices, also known as edge nodes, edge computing is a distributed computing concept that enables real-time data processing and analysis near to the point of data aggregation. Data does not have to be directly uploaded to the cloud or a centralized data processing system with edge computing \cite{ref6}. By adopting edge computing, most businesses use centralized storage, often in a public cloud or private cloud environment, to store, manage, and analyze their data \cite{ref7}. Unfortunately, many real-world applications no longer have the needs that can be satisfied by traditional infrastructure and cloud computing. To analyze enormous amounts of data in real-time, which is not achievable on a typical information technology (IT) infrastructure, IoT and internet of everything (IoE) require a highly accessible network with minimal latency. 

Recently, the use of MEC and energy harvesting (EH) has become increasingly popular in the IoT domain. MEC provides vital functions such as computations, processing and storage capabilities through the edge servers, which are usually placed as near as possible to the mobile user devices, while EH provides a means to prolong the lifetime of IoT devices by continuously supplying energy \cite{ref8,ref9}. However, effective offloading schemes are needed to optimize the performance of MEC-based IoT devices. Edge computing eliminates the need to transport data to the cloud or an on-premises data center for processing and analysis because data is processed near to the source of data gathering \cite{ref10}. The congestion on the networks and servers will be lessened by this strategy. Edge computing is also very useful in the IoT space, especially in industrial IoT (IIoT), due to its faster response time and capacity for real-time data processing \cite{ref11}. Edge computing technology permits additional advances, such as artificial intelligence (e.g., machine learning), in addition to expediting digital transformation for industrial and manufacturing businesses. 
The importance of investigating task scheduling and offloading arises from several aspects, such as the task arrivals possess a heterogeneous nature of dynamics that is experienced by the MEC's nodes as well as the distributed nature of the overall MEC system. Hence, the accumulated workloads of the MEC's edge nodes are more likely to be prone and pass through unbalanced phases, which leads to a high response time and a high resource cost \cite{ref12}. Posterior design is favorable in existing works. According to its concept, it results in high workloads when task offloading occurs. 

Regarding the contribution that non-orthogonal multiple access (NOMA) offers to the process of task offloading in MEC, the non-orthogonal property of NOMA enables it to maintain successful implementation of the MEC and the offloading process of the computation's tasks to MEC servers by optimizing the adopted resource allocation approach. This is very important as NOMA achieves efficient resource usage, which, in turn, is vital for achieving efficient task offloading. According to the idea of superposition coding, different power levels and coding schemes are allocated to different users \cite{ref13}. In the MEC context, MEC servers suffer from computational resource limitations. Therefore, adopting NOMA, firstly, ensures that multiple mobile users could offload their computation tasks simultaneously to the MEC server without enduring severe competition over the available resources. Secondly, NOMA contributes to balancing the allocation of computation resources among mobile users, which is necessary, especially in dense user scenarios \cite{ref14,ref15}. 
   
Regarding the process of MEC task offloading, consider the situation in which a mobile device tends to offload a computation task to an MEC edge server. In this case, the role that NOMA plays would include power and code allocation to ensure that the transmission from the mobile device to the MEC edge server is highly efficient in terms of both energy and spectral efficiency. The latency in any wireless communication system is considerably affected by factors such as the achievable data rate and the overall system reliability over the communication link. Hence, NOMA could be used to optimize the data rate and reliability for each mobile user. As low latency is vital for practical applications in MEC, it is possible to enhance the achievable data rate, which, in turn, reduces the latency for task offloading through the optimization of the allocated power and code rate for each NOMA user. Among the considered techniques, reinforcement learning (RL)-based approaches appear to be a popular method for tackling the challenges of MEC, including the management of limited resources and the optimization of offloading decisions; hence, it inspires the work of this paper to adopt RL approach. Although various mechanisms for optimizing offloading policy in MEC-based IoT devices appeared in open literature, there are still some gaps and areas that future research could investigate.
In light of this, the contributions of this work are as follows:
\begin{itemize}
 \item A new optimized adaptive offloading strategy that allows mobile devices to learn optimal offloading decisions under various scenarios is proposed. A multi-agent deep Q-network is considered to learn and then reach the optimal offloading decision in a distributed manner. 
 \item	The proposed approach outperforms the traditional offloading methods in terms of latency and energy consumption. The proposed scheme selects the edge device and offloading rate based on each server workload. 
 \item	A DRL-based offloading scheme is also proposed to enhance the speed of the learning process. 
 \item	Proposing RL-based mobile offloading scheme for MEC that uses safe RL to avoid adopting an unstable offloading procedure in cases where it fails to achieve the predetermined goals within the system constraints such as the task's computational latency. The proposed strategy permits each mobile device  to select the best edge device, offloading rate and transmit power to optimize its own utility without the necessity of knowledge of the model related to task generation, MEC, or even the interference model.
\end{itemize}

\section{THE PROPOSED SYSTEM MODEL}

\begin{figure*}[!t]
    \centering
    \includegraphics[width=\textwidth]{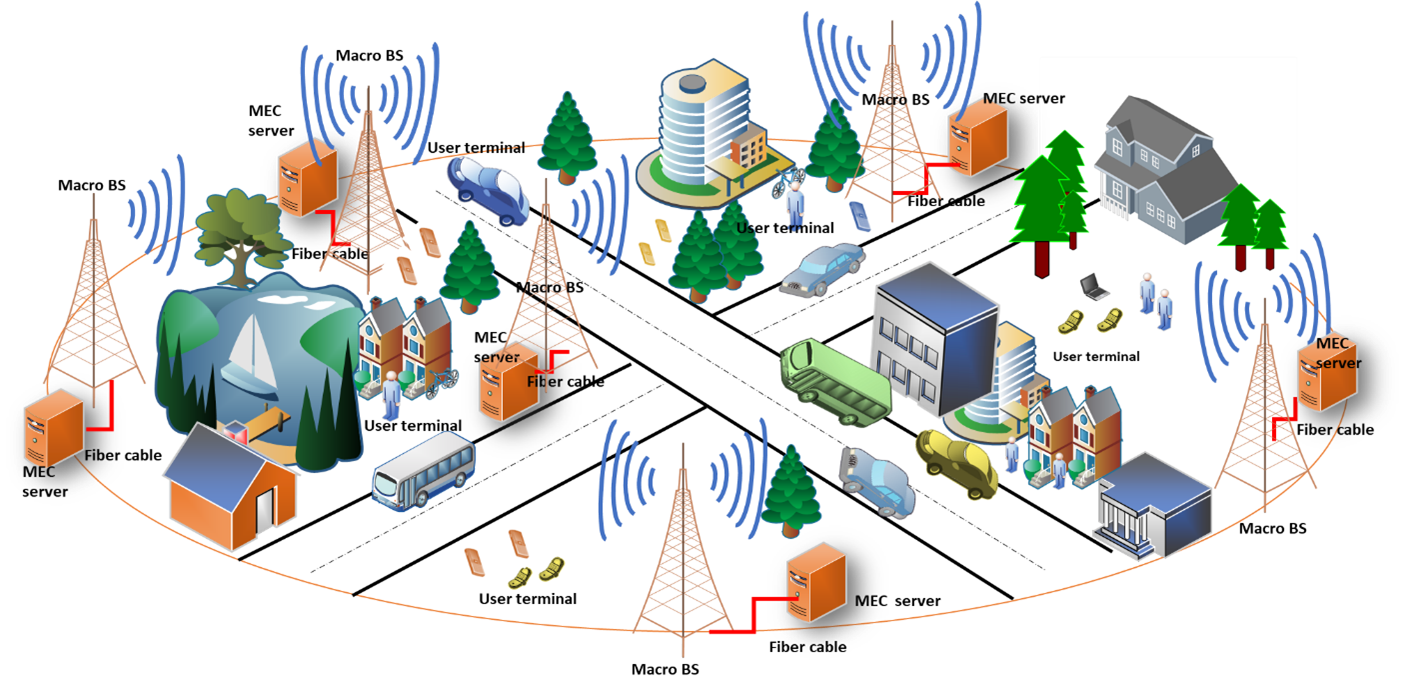}
\caption{An illustration of the considered system model of MEC based IoT enabled ultra dense network.}
        \label{figz}
\end{figure*}

The proposed system model consists of $M$ base stations (BSs) indexed as ${m_1,m_2,m_3,……..m_M}$. The modeled system environment is based on LTE standards that are set up to meet the expected demands of future generations, as shown in Fig. \ref{figz}, with NOMA is chosen as the air interface. It is assumed that at each BS, $m$, there is an edge server, therefore, the delay between the two could be counted as zero. These servers are indexed as ${s_1,s_2,s_3,……..s_M }$. Each edge server is characterized by a computational capacity of a certain limit. The total number of users is set to $K$, with all of which generate a total of T computing tasks.

The computations are modeled so that, for the $i$-th user that has a computation task $\tau_i$, the input data has the size of $D_{i}$. On the other hand, the total CPU cycles that are required for processing each single bit of data are denoted by $S_i$.The offloading decision of each user is modeled as $A = \{ a_{i,j} \}_{K *(M+1)}$, and two computational strategies to be considered in this paper. The case where $a_{(i,j)}=1$ indicates the user task is being offloaded to the BS. On the other hand, the case where $a_{(i,0)}=1$ indicates the task is executed locally. Task execution through offloading is applied when the user sends the computing task for processing to the intended BS. After that, the data computation process is applied by the edge server of that BS. Finally, the BS returns the obtained results to the same user. The duration of time consumed at this stage represents a very small percentage of the transmission time period. Since there is no data transmission at the local execution of the computing tasks, the energy consumed will only be represented by the computing energy, and its value depends on the chip architecture. On the other hand, the energy consumed in the case of offload execution takes into account the energy consumed during the computation phase and the transmission phase. The objectives considered to be maintained by the proposed system model are to minimize both the consumed energy and the time delay during the processing of each task. Achieving such objectives will facilitate maintaining a high degree of Quality of Service (QoS) under the constraint of limited energy in the BS. The first problem has the objective of minimizing the total latency and the overall energy consumption by optimizing the offloading decision for each task. 

The proposed problem could be formulated as a mixed-integer optimization problem using integer-valued weights of the time elapsed during task processing and the energy consumption weight (Assume that they are termed as  $w^{a}$  and  $w^{b}$), respectively. The integer values of those two weights sum up to an integer, constant value termed as $\phi$. For example, if $w^a=w^b$, the offloading within the considered communication scenario assumes that both latency and energy consumption are equally important. Moreover, if $w^a>w^b$,  the considered scenario is assumed to be more sensitive latency. Otherwise, if $w^a<w^b$, the scenario is assumed to be more sensitive to energy consumption. It must be noted that the formulated problem gives insight into the trade-offs between several aspects, i.e., energy consumption, latency, each user capacity, the edge server capacity, and the communication capabilities. By examining and analyzing the results, it is possible to understand how these factors play a vital role in affecting the overall performance of the system and how they contribute to making informed decisions and consequently optimize the system performance. Solving the proposed formulated optimization problem and finding the optimal solution is possible by applying the Mixed Integer Linear Programming (MILP) technique. This is because MILP permits formulating and solving the types of optimization problem that have both continuous and discrete variables, which makes it a good match for the MEC decision-offloading process. 

To exploit the MILP approach, several aspects of the optimization problem need to be predefined, such as the decision variables, the objective function, and the constraints. In this paper, the problem is solved numerically using the MILP solver toolbox in MATLAB, in particular, the intlinprog function is adopted to find the optimal solution. To minimize the overall energy consumed during task execution. The second proposed optimization problem has the objective of minimizing energy consumption for executing the tasks while maintaining the latency constraint. This is important due to the scarcity of power resources and the need to establish an energy-efficient system. The objective function minimizes the total energy consumption for executing all tasks, and the constraints guarantee that the latency experienced by each mobile user for processing each computing task is within the QoS limits.

\section{IMPLEMENTATION OF Q-LEARNING FOR TASK OFFLOADING IN EDGE COMPUTING}
This section presents a task offloading problem that is based on the tabular Q-learning algorithm. The problem setup includes a set of mobile users and a set of BSs, each of which is assumed to be equipped with an edge server. It is assumed that each user has a set of tasks to be processed. These tasks could be executed locally on the user device or offloaded to an edge server. The goal is to find the best offloading decision for each task and each user to minimize the total execution time of the tasks, and hence, minimizing the latency. The Q-learning algorithm is adopted here to assist in learning and make optimal offloading decisions over time. Therefore, the approach in this section is not formulated as a learning problem but rather as an optimization problem. The objective here is to obtain the optimal policy that maximizes the total reward obtained by the agent to make the optimal offloading decision, taking into account the resulting consequences and the acquired rewards. 

\subsection{THE PROPOSED RL-BASED APPROACH}
In each time slot, the user's terminal sends a computing task and those tasks follow a Poisson distribution. Depending on the capacity of the servers and their status, whether being overloaded or not, the computing tasks are processed either by those servers, and this is what is known as local processing, or else the tasks are offloaded. The main goal behind the learning problem is to optimize the task offloading decisions of each mobile user so that their computation time and communication time are minimized. Minimizing the computation time and communication time could directly contribute to maximizing the rewards acquired by those users. To be precise, the proposed decision-making process is applied through sequential steps. The goal of this algorithm is to find the edge server capacity that maximizes the total reward of all users. Firstly, it computes the valid offloading decisions for each mobile user and each BS's edge server. This is done based on the computation and communication times of the tasks and the capacity of edge server capacity. Then the decision-making algorithm picks the next task allocation for each user based on the valid off-loading decisions. Finally, it computes the reward for each user based on the total time required to complete the task.
     
\subsection{THE PROPOSED MODIFIED TABULAR Q-LEARNING ALGORITHM}
The tabular Q-learning algorithm is exploited in this section, which works on updating the Q-value table by mapping the states and actions to their respective rewards. It stores the values in a table where the rows represent the states, and the columns represent the actions. The Q-learning algorithm stores the predicted rewards in its Q-table, which is iteratively updated. These rewards help each agent perform a certain action in a certain state. In order to continuously optimize its own cumulative rewards, each agent is responsible for monitoring its environment and the receipt of rewards. In conventional form, the formula that Q-learning uses for update is given by

\begin{equation}
Q(\phi, \vartheta) \leftarrow Q(\phi, \vartheta) + \delta \left[ r + \beta \max_{a} Q(\phi', \vartheta) - Q(\phi, \vartheta) \right]
\tag{1}
\label{eq2}
\end{equation}
where $Q(\phi, \vartheta)$ is the value of Q at the $\phi$-th state and the $\vartheta$-th action, $\phi$ is the current state, $\vartheta$ is the action taken by the agent, $r$ is the reward acquired by the agent after taking the $\vartheta$-th action in the $\phi$-th state,  $\beta$ is the discount factor reflecting the importance of future rewards and $\phi'$ is the next state reached after the agent takes the $\vartheta$-th action.

It should be noted that Q-learning sometimes acts less efficiently in environments characterized with large state-spaces. In this paper, a modification is proposed to produce a more advanced strategy in terms of exploitation and exploration. It is also assumed that the agents adopt the concept of the $\varepsilon$-greedy scheme. Each agent randomly monitors the premises with an exploration probability of $\varepsilon$ and performs the best possible action with a probability of $(1-\varepsilon)$. Hence, after applying the proposed modification to (\ref{eq2}), the updated equation of Q-learning approach will be given as
\begin{align}
Q(\phi, \vartheta) \leftarrow Q(\phi, \vartheta) 
&+ \delta \bigg[ r + \beta \Big( (1 - \epsilon)
\notag
\\
&\, \times \max_{a} Q(\phi', \vartheta) \nonumber + \epsilon \tau \Big) - Q(\phi, \vartheta) \bigg]
\tag{2}
\label{eq3}
\end{align}
where $\tau$ represents a random exploration weight that is also affected by the difference between the current reward value and the reward value previously assigned to the same agent. In such a case, the proposed rewarding strategy will encourage the agent exploration to be focused on a direction that almost guarantees benefit in favor of the Q-value and hence, the whole system. After that, each agent updates the Q-values for each state-action pair based on the proposed approach in (\ref{eq3}), which has a type of recursive formula that relates the value of a state to the values of its closest states. After applying (\ref{eq3}), the reward value $r$ is also stored to be checked after each action in such a way that if the acquired reward is worth the action taken by the agent, then the agent would proceed in its action. However, if the newly obtained state is not much different from the previous one, the reward would be set at the same value to encourage the agent to take better action. This is very important in dense scenarios such as ultra-dense IoT networks considered in this paper. The proposed RL-based algorithm then performs the selected actions and receives a reward for each end-node. 

\subsection{THE PROPOSED REWARD CALCULATIONS}
In the proposed RL-based algorithm, the reward is simply calculated as the negative quantity of the total execution time (total latency) for each user's tasks. In this case and to obtain a precise result, it will include both the computation times and communication times. The proposed learning algorithm then updates the Q-table according to the received rewards and the expected rewards for the next state and action. The Q value table is updated using a learning rate, a discount factor, and the received and expected rewards. It is worth mentioning that the algorithm repeats this process for a fixed number of episodes.  After applying the proposed modified tabular Q-learning algorithm within the proposed approach, the algorithm determines the offloading decision surface, taking into consideration the system model environment, including the dense nature of the network. This is to show the predicted total reward for each offloading task and each corresponding edge server. Then, the proposed algorithm counts the average reward received by every end-node over all episodes and all the Monte Carlo simulation iterations. 

\section{SIMULATIONS AND RESULTS}
The simulated scenario emulates a dense situation where around 10 to 100 nodes are active in the covered area, and each one is assumed to have up to 10 tasks. The total number of servers is chosen to be 5. The communication speed is set as 10Mb/s while the speed of processing is set to around 1Gb/s and the data size is 1Gb/s. The carrier frequency of this setup is set to 5GHz. The non-uniform allocation of the MEC serves and the nodes are adopted randomly for a more practical scenario.

\begin{figure}[t]
    \centering
    \includegraphics[trim=0 0 0 20, clip, width=0.45\textwidth]{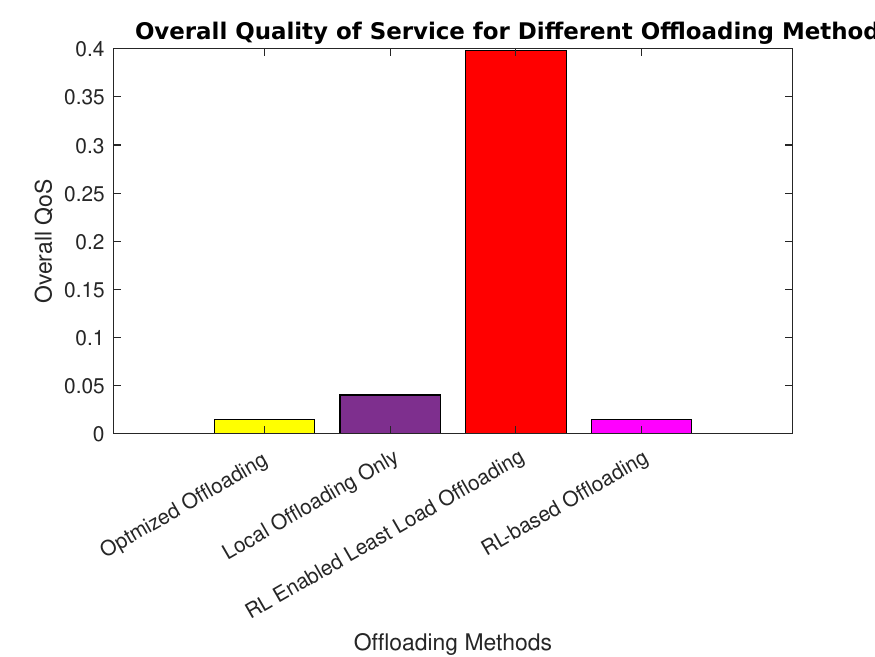}
    \caption{Comparison of the proposed offloading schemes in terms of the level of QoS provided by each approach.}
    \label{fig10_1}
\end{figure}

Fig. \ref{fig10_1} shows the QoS performance for the optimized off-loading, the local offloading approach that presumes to perform all the offloading locally, the proposed RL enabled least load offloading, and the proposed RL enabled offloading scheme. Note that the QoS is calculated by taking into consideration the energy consumption and latency for each scheme. 

\begin{figure}[t]
    \centering
    \includegraphics[trim=0 0 0 20, clip,width=0.45\textwidth]{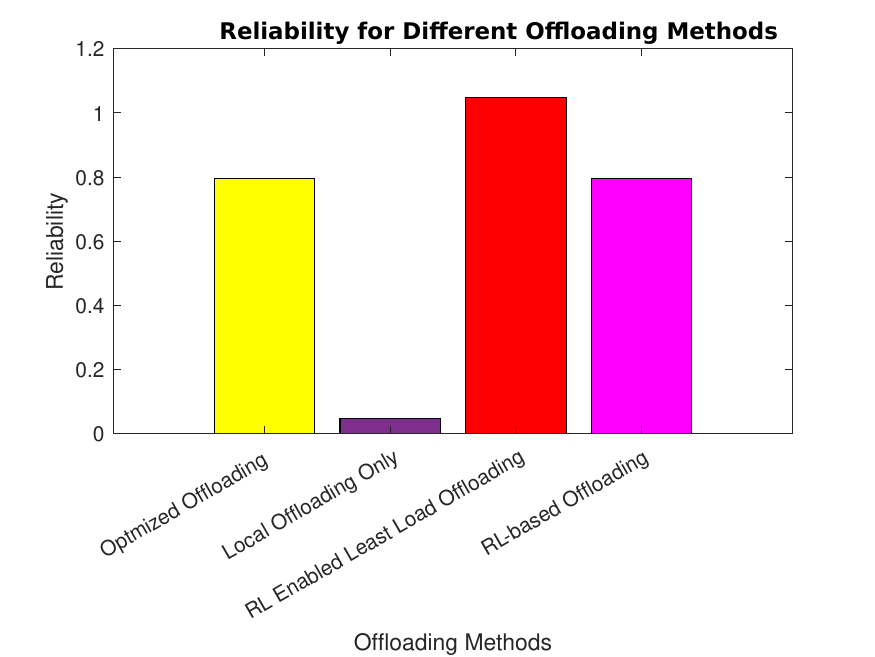}
    \caption{Comparison among the proposed offloading schemes in terms of the level of reliability.}
    \label{fig11_2}
\end{figure}

The reliability of the compared schemes is examined in Fig. \ref{fig11_2}. The proposed optimized approach and the proposed RL based offloading approach both offer comparable levels of reliability but lower than that of the proposed RL-enabled least load offloading scheme. All three of those schemes offer better reliability than the local execution approach. Fig. \ref{fig11_2} also shows that the offloading scheme that depends on RL and chooses the server with the least load to perform the offloading offers the best QoS among all the schemes. This is mainly because of the attributes of the proposed RL that benefit from earlier learning of all servers' conditions before deciding on the offloading server. This leads to the efficient usage of energy and less latency spent in executing the offloading process.

\begin{figure}[t]
    \centering
    \includegraphics[trim=0 0 0 20, clip,width=0.45\textwidth]{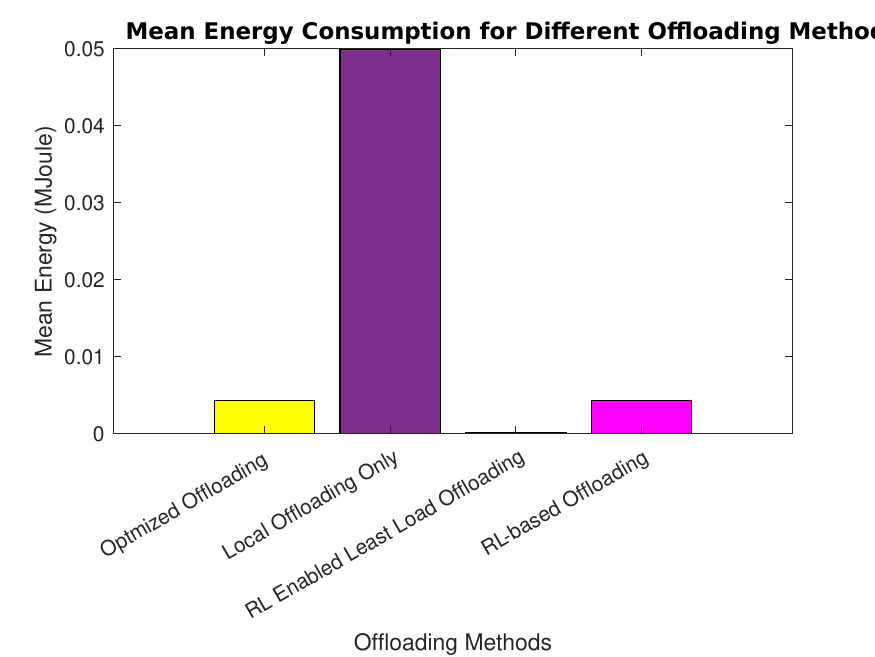}
    \caption{Comparison among the proposed offloading schemes in terms of the average energy consumed by each approach.}
    \label{fig12_3}
\end{figure}

The average energy consumption is shown in Fig. \ref{fig12_3} for all the proposed schemes. This figure illustrates that the RL enabled least offloading scheme consumes the least energy while offloading a task compared to the other schemes. This highlights the vital role of RL in choosing the offloading server, as well as the importance of selecting suitable rewarding and learning approaches.

\begin{figure}[t]
    \centering
    \includegraphics[trim=0 0 0 20, clip,width=0.45\textwidth]{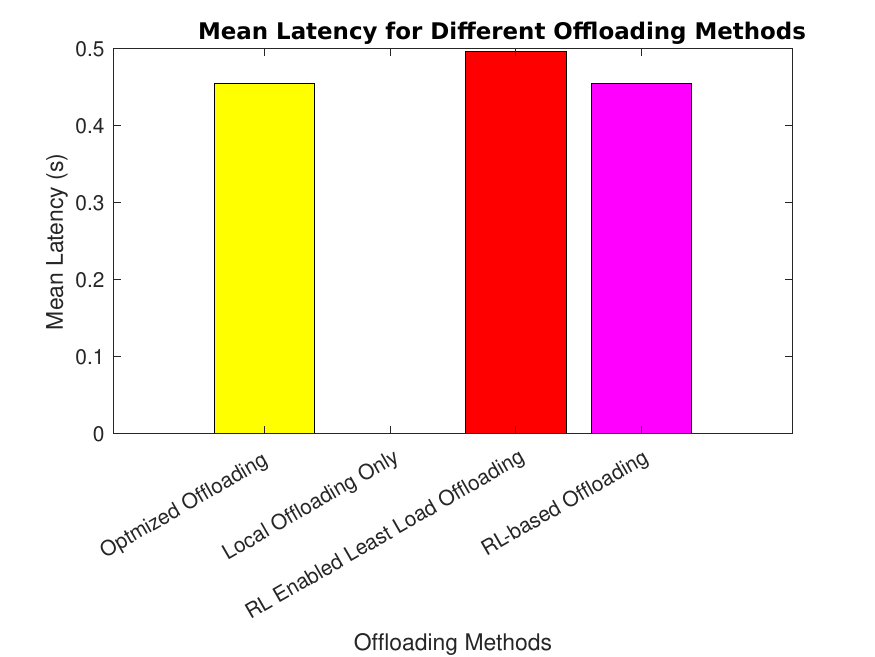}
    \caption{Comparison of the proposed offloading schemes in terms of the average latency consumed by each approach.}
    \label{fig13_4}
\end{figure}

Fig. \ref{fig13_4} shows a comparison of the average latency of each of the proposed schemes against the local execution of the task offloading process. Examining the latency of each offloading approach is important, especially for 6G-enabled IoT network applications. Predictably, the local execution scheme clearly offers the least latency given that there is no delay or latency as all related operations are performed locally. On the other hand, the proposed RL-enabled least load offloading strategy offers a slightly higher latency than the other two proposed schemes. This slight latency part could be a result of the learning and server-selecting process which could be lessened by selecting another learning or rewarding technique. All in all, the proposed RL enables least load offloading scheme offers the best performance as compared to the proposed RL-based offloading and optimized offloading, with slightly higher latency. On the other hand, the proposed optimized offloading strategy shows promising behaviors in terms of energy consumption, latency, QoS and reliability but still falls behind the proposed RL enabled load offloading scheme. Choosing the right offloading scheme might well depend upon the given application by taking into account energy consumption and latency as priorities. 

\section{Future Research Directions}
 While the proposed offloading framework using optimization and RL demonstrates promising performance for MEC in wireless IoT networks, there are still a number of areas that need further research and development.  In addition to improving system performance, these efforts seek to increase the robustness and practical application of MEC-based offloading solutions under actual constraints.

\subsection{Federated Learning Integration for Privacy-Preserving Offloading}
 Integrating federated learning (FL) with offloading choices is a crucial future path.  Centralized training in existing RL-based offloading systems may jeopardize user privacy, particularly in delicate IoT domains such as industrial monitoring or healthcare.  Without exchanging raw data, FL enables edge devices to work together to train global models.  In accordance with data protection laws such as the GDPR, integrating FL into the MEC offloading process would preserve learning effectiveness while protecting privacy.
 
\subsection{Sustainable and Energy-Efficient MEC Offloading}
 Future offloading systems must prioritize sustainability and energy economy, even though performance measures such as latency and throughput have been highlighted.  Typically, IoT devices have limited resources, and edge servers themselves add to the total power consumption of the network.  Green offloading algorithms that reduce energy consumption while preserving a satisfactory QoS should be the focus of future research.  Multi-objective optimization approaches that balance latency, throughput, and energy usage in offloading decisions could help achieve this.

\subsection{Using Meta-RL to Maintain Quick-Adaptation}
 Conventional RL agents may perform worse in dynamic wireless contexts because of user mobility or non-stationary network conditions.  Using meta-RL, which enables agents to quickly adjust to novel tasks or surroundings with little data, is one possible approach.  This could greatly increase the system's resilience in practical deployments by improving its responsiveness to erratic changes in network topology, load variations, or shifting user demands.
 
\subsection{Joint Optimization of Computation, Communication, and Caching}
Our approach, similar to the majority of existing solutions, concentrates only on compute offloading.  On the other hand, a more comprehensive strategy would optimize caching, communication, and computing all at once.  By prefetching or storing popular material at edge nodes, such a collaborative design could minimize unnecessary computations and data transmissions.  Deep multi-agent RL frameworks that cooperatively manage dispersed resources or sophisticated graph-based RL techniques could be used to mimic this.

\subsection{Deployment and Scalability in Diverse Environments}
 A wide range of devices, edge servers, and communication protocols are used in real-world MEC systems, which are extremely diverse.  In order to solve scalability and interoperability concerns, future studies should create offloading models that are lightweight and flexible enough to function well on a variety of platforms.  To handle such complexity, cloud-edge-device collaborative frameworks and hierarchical reinforcement learning might be crucial.

\subsection{Offloading Techniques That Consider Security}
 Cyberthreats such as data manipulation, denial-of-service assaults, and hostile RL model manipulation can affect IoT systems.  The creation of security-aware offloading systems that recognize and lessen such assaults would be a useful addition.  To increase system trustworthiness, this could entail safe multi-party computation methods, blockchain-based authentication for MEC nodes, or adversarial training.

\section{CONCLUSIONS}
This paper discussed how the use of RL-based adaptive task offloading in IoT networks and MEC-assisted ultra-dense networks can offer a promising solution for efficient and intelligent resource management in IoT networks and applications. This paper also proved that leveraging the features of machine learning algorithms enables dynamic and adaptive task allocation, which contributes to improving the overall network performance, as well as maintaining reduced latency and enhancing the overall network scalability. Moreover, exploiting NOMA boosts the successful implementation of MEC and task offloading through power optimization and efficient resource allocation. It also works on reducing the overall latency and efficiently managing the scarcity of computational resources. It is clear from the results obtained that there is proportion between the compared entities. This is mainly because having more mobile users means more tasks to offload, and more energy to be consumed in the offloading process. In addition, a higher number of mobile users means there will be increased competition among those users over the available resources. All in all, due to the high volume of data created and the distance between clients and the backend data centers, typical centralized cloud computing would be concerning for dense networks with limited resources and a large number of IoT devices. As MEC distributes computing, control, storage, and networking services closer to nodes, it can effectively address many of the constraints of the cloud-only approach.

\end{document}